\documentclass[nofootinbib,aps,prd,nobibnotes,twocolumn,superscriptaddress,bibliography,10pt]{revtex4-2}

\usepackage{amsfonts}
\usepackage{mathrsfs}
\usepackage{amsmath}
\usepackage{color}
\usepackage{graphicx}
\usepackage{bm}
\usepackage{amssymb}
\usepackage{xspace}
\usepackage{epstopdf}
\usepackage{multirow}
\usepackage{braket}
\usepackage[colorlinks=true, letterpaper=true, pdfstartview=FitV, linkcolor=blue, citecolor=blue, urlcolor=blue]{hyperref}
\usepackage{soul}
\usepackage{xcolor}
\usepackage{tabularx}

\usepackage[normalem]{ulem}

\def \Z {\mathbb{Z}}
\def \M {\mathcal{M}}
\begin{document}
\title{Projective crystal symmetry and topological phases}
\author{Chen Zhang}

\affiliation{Department of Physics and HK Institute of Quantum Science \& Technology, The University of Hong Kong, Pokfulam Road, Hong Kong, China}

\author{Shengyuan A. Yang}
\affiliation{Research Laboratory for Quantum Materials, Department of Applied Physics, The Hong Kong Polytechnic University,
Kowloon, Hong Kong, China}

\author{Y. X. Zhao}
\email[]{yuxinphy@hku.hk}
\affiliation{Department of Physics and HK Institute of Quantum Science \& Technology, The University of Hong Kong, Pokfulam Road, Hong Kong, China}

\begin{abstract}
Quantum states naturally represent symmetry groups, though often in a projective sense.
Intriguingly, the projective nature of crystalline symmetries has remained underexplored until very recently. A series of groundbreaking theoretical and experimental studies have now brought this to light, demonstrating that projective representations of crystal symmetries lead to remarkable consequences in condensed matter physics and various artificial crystals, particularly in their connection to topological phenomena. In this article, we explain the basic ideas and notions underpinning these recent developments and share our perspective on this emerging research area. We specifically highlight that the appearance of momentum-space nonsymmorphic symmetry is a unique feature of projective crystal symmetry representations. This, in turn, has the profound consequence of reducing the fundamental domain of momentum space to all possible flat compact manifolds, which include torus and Klein bottle in 2D and the ten platycosms in 3D, presenting a significantly richer landscape for topological structures than conventional settings. Finally, the ongoing efforts and promising future research directions are discussed.

\end{abstract}
\maketitle

\section{Introduction}
Symmetry groups are generally represented projectively, a concept first highlighted by Wigner almost a century ago~\cite{wigner1939unitary}.
Although crystalline symmetries played a fundamental role in the development of group theory, condensed matter physics, and more recently the topological states of matter, for a long period of time, only ordinary representations of crystal symmetry groups
were considered. Recently, it was realized that the phase degrees of freedom (of a wave function)
can be mixed with spatial transformations, for instance under gauge fields, in both condensed matter and various artificial crystal systems~\cite{wen2002quantum,cualuguaru2024new,xiao2024spin,zhao2020z,shao2021gauge,PhysRevLett.126.196402},
for which one has to consider projective representations of crystal groups, and the resulting projective crystal symmetry algebras may generate totally unexpected remarkable physical consequences~\cite{PhysRevLett.126.196402,Fan_Zhang_PRL,xue2022projectively,li2022acoustic,meng2022spinful,PhysRevLett.132.236401,PhysRevLett.132.213801,PhysRevLett.130.236601,xiao2024revealing,zhao2020z,shao2021gauge,PhysRevB.108.205126}.


Let's first explain what is a projective representation. Consider a crystal group $G$ with elements denoted as $g_i$. Recall that an ordinary representation of $G$ is a mapping $\rho$ of each $g_i\in G$ to an invertible linear operator $\rho(g_i)$ on a Hilbert space $V$, which satisfies the homomorphism property, i.e.,
\begin{equation}\label{ord}
    \rho(g_1)\rho(g_2)=\rho(g_1 g_2),
\end{equation}
for any pair of elements $g_1,g_2\in G$.
By choosing a basis of $V$, each $\rho(g_i)$ can be casted into a matrix form. Now, a projective representation is a generalization of the above concept. Quantum states reside in projective Hilbert space, where states differing by a phase factor are physically indistinguishable. This phase ambiguity means that the group algebra of the symmetry operators can be modified by phase factors, i.e., the symmetry operators projectively represent the group algebra. Specifically, the projective representation is characterized by a multiplier (or factor system) $\nu: G\times G\rightarrow U(1)$, and the homomorphism condition Eq.~(\ref{ord}) is modified as
\begin{equation}\label{eq:p_multi}
    \rho(g_1)\rho(g_2)=\nu(g_1,g_2)\rho(g_1g_2).
\end{equation}
Here, the difference is just the insertion of  a $U(1)$ phase $\nu(g_1,g_2)$ for each pair $g_1,g_2\in G$.
We refer to this $\rho$ as a $\nu$-representation of $G$.

One can see that if for all pairs $(g_1,g_2)$, $\nu=1$ (the trivial case), then it reduces back to the ordinary representation.
For a function $\nu$ to form a valid factor system, it has to satisfy a certain consistency condition, known as
2-cocycle equations (see Eq.~(\ref{eq:self-consistency}) below). Clearly, a nontrivial multiplier $\nu$
modifies the algebraic relations of crystal symmetries and thus enriches topological structures beyond the framework of ordinarily represented crystal symmetry.

Pioneering examples include the M{\"o}bius insulator protected by projective translation symmetries~\cite{zhao2020z}  and the Stiefel-Whitney nodal line semimetal enabled by projective symmetry algebras~\cite{shao2021gauge}. Interestingly, it was discovered that in the framework of projective symmetry, spinless and spinful topological classifications may be switched~\cite{PhysRevLett.126.196402}, namely, with the help of projective symmetry, some topological states which originally only exist in spinless systems can now be realized in spinful systems, and vice versa.
For example, this idea leads to spinless mirror Chern insulator, and its analogue has also been studied in non-Hermitian systems~\cite{PhysRevB.108.205126,wang2025non,rui2023making}.



\begin{figure}[!t]
    \centering
    \includegraphics[width=\linewidth]{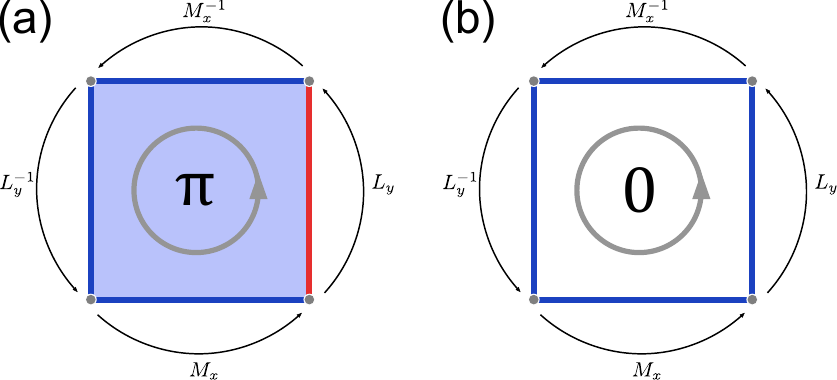}
    \caption{(a) and (b) illustrate rectangular plaquettes with fluxes $\pi$ and $0$, respectively. Negative and positive hopping amplitudes are marked in red and blue, respectively. Successive operations $L_y^{-1}M_x^{-1}L_yM_x$ move a particle around a rectangle, leading to the nontrivial projective algebra relation if $\phi=\pi$.}
    \label{fig_flux}
\end{figure}

How to realize a nontrivial $\nu$? In crystals, a nontrivial $\nu$ is associated with an appropriate static gauge flux configuration. To understand this, consider the simple example illustrated in Fig.~\ref{fig_flux}.
Focus on two crystal symmetries: lattice translation symmetry $L_y$ and mirror symmetry and $M_x$. The two operations commute, $L_y M_x=M_x L_y$.
Consider their representation in the Hilbert space corresponding to a particle hopping around the four sites of the rectangle. If there is no gauge flux through the rectangle, we should have $\rho(L_y)^\dagger \rho(M_x)^\dagger\rho(L_y)\rho(M_x)=1$, and this is just the ordinary representation. On the other hand, if
there is a gauge flux $\phi$ through the rectangle, the relation is modified to
\begin{equation}\label{flux}
\rho(L_y)^\dagger \rho(M_x)^\dagger \rho(L_y)\rho(M_x)=e^{i\phi},
\end{equation}
since hopping around the rectangle gives an extra $\phi$ phase due to the flux. Clearly, the phase $e^{i\phi}$ is invariant under arbitrarily modifying the phase of the unitary operator for each symmetry. Moreover, the phase can only be $0$ or $\pi$, which can be shown by first converting Eq.~\eqref{flux} into $ \rho(L_y)\rho(M_x)=e^{i\phi} \rho(M_x)\rho(L_y)$ and then multiplying both sides with $\rho(M_x)$, noting that $\rho(M_x)^2=\pm 1$.  This is also consistent with the fact that only fluxes $0$ and $\pi$ can preserve the mirror symmetry reflecting the plaquette. One can easily see that Eq.~\eqref{flux}
corresponds to a nontrivial multiplier $\nu$ with
\begin{equation}
  \nu(M_x,L_y)=e^{i\phi},\qquad \nu(L_y,M_x)=1.
\end{equation}

A common scenario to have such flux is for solid crystals under a magnetic field. However, since the lattice constants for most materials are small (on the order of angstroms), to have appreciable $\phi$, the required magnetic field strength is usually huge. Meanwhile, the rapid development of various artificial crystals, such as acoustic/photonic crystals, electric circuit arrays, and mechanical metamaterials, has made it very convenient to engineer gauge flux configurations in crystals~\cite{lu2014topological,huber2016topological,yang2015topological,ozawa2019topological, mittal2019photonic,Xue_2020, xue2021observation,ma2019topological,zhang2018topological, cooper2019topological, zhu2023topological,PhysRevA.110.023308,dalibard2011colloquium,prodan2009topological,imhof2018topolectrical, yu20204d, wang2023realization}.
Indeed, several predictions on projective crystal symmetries (an abbreviated name for projective representation of crystal symmetry) have been successfully verified using artificial crystals~\cite{li2022acoustic,xue2022projectively,yue2023acoustic,meng2022spinful,li2023acoustic,liu2024topological,liu2023mobius,jiang2023photonic,liu2024topological,PhysRevB.109.134107,PhysRevB.108.L220101,zhu2024brillouin,PhysRevApplied.21.044002}

A unique feature of projective crystal symmetry is that it may contain operations that is nonsymmorphic in momentum space.
Previously, `nonsymmorphic' is a character defined only in real space, i.e., a rotation or a mirror operation followed by a
fractional lattice translation in real space (known as screw axis and glide mirror, respectively). In momentum space, all crystal symmetry operations are symmorphic.
However, it was discovered that this conclusion only holds for ordinary representations. When projectively represented,  we may also have momentum-space screw axis and glide mirror symmetries~\cite{chen2022brillouin,PhysRevLett.130.256601}. We will explain this remarkable feature in more detail below.

\begin{figure}[!t]
    \centering
    \includegraphics[width=\linewidth]{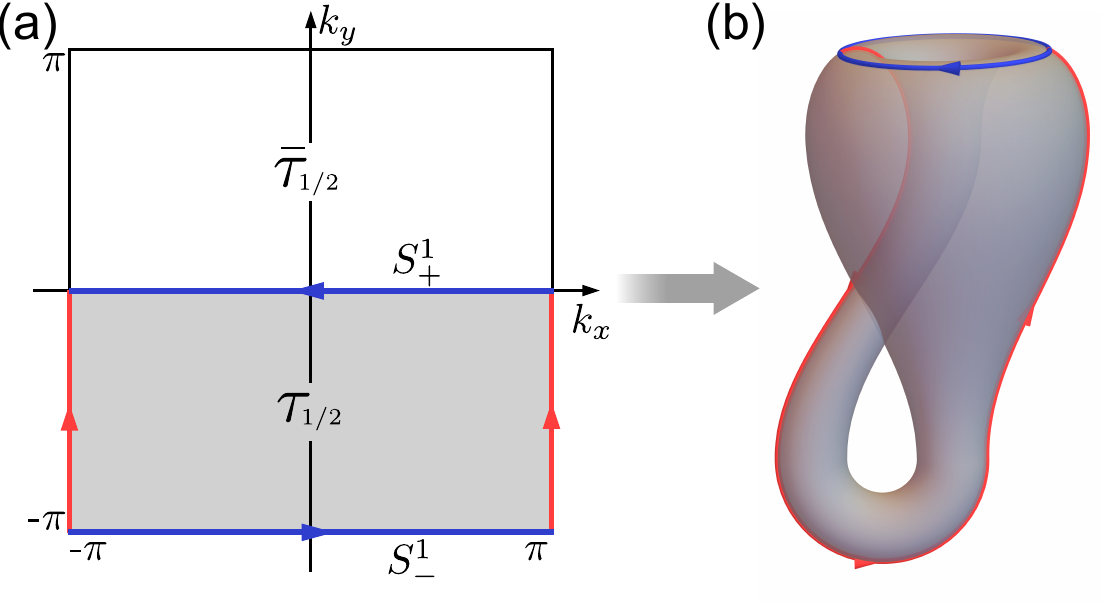}
    \caption{Momentum space representation of Brillouin Klein bottle. (a) The fundamental domain of the Brillouin zone is $\tau_{1/2}$.  The boundaries with the same color should be identified along the marked direction. (b) The fundamental domain $\tau_{1/2}$ with boundaries identified is the Brillouin Klein bottle.}
    \label{fig_klein}
\end{figure}

The emergence of momentum space nonsymmorphic symmetry has profound consequences. Most notably, it may completely change the topology of the elementary unit of momentum space. For ordinary cases, the elementary unit, formally known as the fundamental domain, is our familiar Brillouin zone, having the topology of a torus. In 2D, it is the 2-torus $T^2$.
It was found that with projective crystal symmetry that is nonsymmorphic in momentum space, the fundamental domain may be transformed to a Klein bottle $K^2$~\cite{chen2022brillouin} (see Fig. ~\ref{fig_klein}), which is a non-orientable object. This change also necessitates a change in the
topological classifications of physical states. For example, a general band insulator on the Brillouin Klein bottle has a $\mathbb{Z}_2$ classification, instead of $\mathbb{Z}$. This discussion can be extended to higher dimensions.
In 3D, with projective crystal symmetry, one can realize fundamental domains having topology of all possible flat compact manifolds, known as the ten platycosms. And each Brillouin platycosm has its special topological classification of band structures.

The framework of projective crystal symmetry can be further extended by internal symmetries in the tenfold Altland-Zirnbauer symmetry classes, including time-reversal, particle-hole and chiral symmetries~\cite{altland1997nonstandard,heinzner2005symmetry}. Some efforts have been made along this line. For example, the anti-unitary time-reversal symmetry $\mathcal{T}$ can constrain the factor systems to be valued in $\mathbb{Z}_2$, and projective representations of wallpaper groups in the presence of $\mathcal{T}$ have been classified by formulating a complete set of cohomology invariants~\cite{chen2023classification}. The pioneering work Ref.~\cite{zhao2020z} proposes the M\"{o}bius insulator in the presence of chiral (sublattice) symmetry. Recent studies further examine the symmetry algebra between lattice translation and sublattice symmetry, leading to topological classifications distinct from the conventional case~\cite{xiao2024revealing}. The aforementioned switch of spinful and spinless topological classifications also represents a new result in this extended framework~\cite{PhysRevB.108.205126,PhysRevLett.126.196402}.

In the following, we shall first give a closer look at projective representations of crystal symmetry groups, highlighting the emergence of momentum-space nonsymmorphic character. Then, we explain how the nonsymmorphic character leads to change in topology of the fundamental domain of momentum space and how the change necessitates a change in topological classification of band structures. We further discuss how the framework can be refined by including internal symmetries. Finally, we discuss on-going efforts and promising future directions in this newly emerging research area.

\section{Projective crystal symmetry}

Let's examine the projective representation of crystal symmetry in more detail. For simplicity, we shall
consider a symmorphic space group $G$ in all the discussions below.
It contains two natural subgroups, namely, the translation group $L$ and the point group $P$. $L$ consists of translations with lattice vectors $\bm{t}=\sum_in^i\bm{e}_i$, where $n^i$ are integers and $\bm{e}_i$ are the primitive lattice vectors. We may write
\begin{equation}
    L=\{\bm{t}=\sum_i n^i\bm{e}_i,n^i\in\mathbb{Z}\}.
\end{equation}
$L$ can also be interpreted as the collection of all lattice sites related to the origin by lattice translations. Meanwhile, the point group $P$ is a finite subgroup of the orthogonal group $O(d)$, which operates on the lattice $L$ in a compatible manner, i.e., $R\bm{t}\in L$ for any $R\in P$ and $t\in L$. Such a compatible pair $(P,L)$ is referred to as a crystal class $D$. For the symmorphic $G$ considered here, we have $G=L\rtimes_D P$, having a simple structure of a semidirect product of $L$ and $P$. Indeed, any element of $G$ can be expressed as
$(\bm{t},R)$ with $\bm{t}\in L$ and $R\in P$, and the group multiplication is given by
\begin{equation}
    (\bm{t}_1,R_1)(\bm{t}_2,R_2)=(\bm{t}_1+R_1\bm{t}_2,R_1R_2).
\end{equation}

Consider a projective representation $\rho$ of $G$ with factor system $\nu$. The associativity condition of group multiplication
requires  $\nu$ to satisfy the following self-consistency equation
\begin{equation}\label{eq:self-consistency}
    \nu(g_1,g_2)\nu(g_1g_2,g_3)=\nu(g_1,g_2g_3)\nu(g_2,g_3),
\end{equation}
for all $g_1,g_2,g_3\in G$. This is also known as the 2-cocycle equation.

Note that there is certain redundancy in the choice of $\nu$ (just like a gauge degree of freedom). If we transform $\rho(g)$ to $\rho^{\prime}=\chi(g)\rho(g)$ using a phase factor $\chi(g)\in  U(1)$, then the multiplier $\nu(g_1,g_2)$ is transformed as
\begin{equation}\label{gauge}
\nu^{\prime}(g_1,g_2)=\nu(g_1,g_2)\chi(g_1)\chi(g_2)/\chi(g_1g_2).
\end{equation}
The extra factor $\chi(g_1)\chi(g_2)/\chi(g_1g_2)$ in Eq.~(\ref{gauge}) is known as a coboundary. Clearly, $\nu$ and $\nu'$ describe the same underlying algebraic structure, hence, two multipliers that differ from each other by a coboundary are considered equivalent. Mathematicians have developed a way to classify all inequivalent factor systems, by using so-called second group cohomology $H^2(G,U(1))$. This approach also provides a set of cohomology invariants for characterizing factor systems.

Particularly, for symmorphic space groups, their multiplier $\nu$ admits a simple canonical decomposition through Mackey's formula~\cite{mackey1958unitary,mackey1989unitary}:
\begin{equation}\label{eq:mackey}
\begin{split}
    \nu[(\bm{t}_1,R_1),(\bm{t}_2,R_2)] =&\ \sigma(\bm{t}_1,R_1\bm{t}_2) \cdot \\
    &\ \gamma(R_1\bm{t}_2,R_1)\alpha(R_1,R_2),
\end{split}
\end{equation}
where $\sigma$ and $\alpha$ represent the restrictions of $\nu$ onto the translation subgroup $L$ and the point group $P$, respectively, while $\gamma$ encodes phase factors induced by the action of $P$ on $L$.

Non-trivial factors of translation and point groups frequently emerge in condensed matter physics. Magnetic translation operators $\mathcal{L}_i$, characterized by modified phase factors from magnetic flux $\phi=2\pi p/q$, are in fact projective representations of the translation group~\cite{zak1964magnetic}. Their non-commutativity $\mathcal L_x \mathcal L_y\mathcal L_x^{-1}\mathcal L_y^{-1} = e^{i2\pi p/q}$ requires enlarging the unit cell, yielding the magnetic Brillouin zone, which is crucial for understanding the quantum Hall effect~\cite{thouless1982quantized,cage2012quantum}. Meanwhile, projective algebras of point groups often arise in spinful systems, e.g., as exemplified by the mirror symmetry relation $\mathcal M^2=-1$ in mirror Chern insulators~\cite{teo2008surface,hsieh2012topological,tanaka2012experimental}.

In Eq.~(\ref{eq:mackey}), the $\gamma$-component actually governs the modified action of point-group symmetries in momentum space. It is the key for the emergence of nonsymmorphic symmetry in momentum space~\cite{PhysRevLett.130.256601}.
To see this, we consider a factor system $\nu$ of $G$, whose restriction to both $L$ and $P$ are trivial, i.e., $\sigma=\alpha=1$ in Eq.~(\ref{eq:mackey}). Then the multiplier only depends on $\gamma$,
%
\begin{equation}\label{gamma}
    \nu[(\bm{t}_1,R_1),(\bm{t}_2,R_2)]=\gamma(R_1\bm{t}_2,R_1).
\end{equation}
Here, $\gamma(\bm{t},R)$ is valued in $U(1)$, and it is easy to show that $\gamma$ must satisfy the following conditions
\begin{equation} \label{11}
\gamma(\bm{t}_1+\bm{t}_2,R)=\gamma(\bm{t}_1,R)\gamma(\bm{t}_2,R),
\end{equation}
and
\begin{equation}\label{eq:gamma}
\gamma(\bm{t},R_1R_2)=\gamma(\bm{t},R_1)\gamma(R_1^{-1}\bm{t},R_2).
\end{equation}

These two conditions ensure the multiplier $\nu$ satisfies the requirement of Eq.~\eqref{eq:self-consistency}. Equation (\ref{11}) implies $\gamma(*,R)$ with $R$ fixed is a homomorphism from $L$ to $U(1)$, hence, $\gamma$ should take the following form:
\begin{equation}\label{eq:gamma2}
    \gamma(\bm{t},R)=e^{-i\bm{\kappa}_R\cdot \bm{t}},
\end{equation}
where $\bm{\kappa}_R$ is a momentum space vector specifying the homomorphism for each $R\in P$. Therefore, from Eq.~(\ref{gamma}), the multiplier can be expressed as
\begin{equation}\label{eq:multiplier}
    \nu[(\bm{t}_1,R_1),(\bm{t}_2,R_2)]=e^{-i\bm{\kappa}_{R_1}\cdot R_1\bm{t}_2}.
\end{equation}
Evidently, $\bm{\kappa}_R$ depends on $R$. Substituting Eq.~\eqref{eq:gamma2} into Eq.~\eqref{eq:gamma} gives the following constraint on $\bm{\kappa}_R$:
\begin{equation}\label{eq:kappa}
    \bm{\kappa}_{R_1}+R_1\bm{\kappa}_{R_2}-\bm{\kappa}_{R_1R_2}\in \hat{L},
\end{equation}
for any $R_1,R_2\in P$. Here, $\hat{L}$ is the reciprocal lattice, i.e., from $\bm{e}_i$, one can derive the primitive reciprocal lattice vectors $\bm{G}_i$, then $\hat{L}$ is given by
\begin{equation}
    \hat{L}=\{\bm{K}=\sum_im^i\bm{G}_i,m^i\in\mathbb{Z}\}.
\end{equation}

We shall show that a nontrivial $\bm{\kappa}_R$ corresponds to the fractional reciprocal lattice translation associated to $R$ operation. In other words, in momentum space, $R$ transforms a momentum vector $\bm{k}$ as
\begin{equation}\label{eq:R_action}
    R:\bm{k}\mapsto R\bm{k}+\bm{\kappa}_R.
\end{equation}

Recall that each momentum $\bm{k}$ labels an irreducible representation $\rho_{\bm{k}}$ of the translation subgroup $L$, with $\rho_{\bm{k}}(\bm{t},1)=e^{i\bm{k}\cdot\bm{t}}$. Then, the action of $(\bm{t}^{\prime},R)$ on $\bm{k}$
is defined as
\begin{equation}
  \rho_{\bm{k}}(\bm{t},1)\mapsto [\rho(\bm{t}',R)]^\dagger\rho_{\bm{k}}(\bm{t},1)\rho(\bm{t}',R).
\end{equation}
The projective algebraic relation implies that
\begin{equation}\label{eq:Conjugation}
	[\rho(\bm{t}',R)]^\dagger\rho(\bm{t},1)=e^{-i\bm{\kappa}_{R^{-1}}\cdot R^{-1}\bm{t}}\rho(R\bm{t},1)[\rho(\bm{t}',R)]^\dagger.
\end{equation}
This leads to
\begin{equation}\label{21}
	e^{i\bm{k}\cdot\bm{t}}\mapsto e^{-i\bm{\kappa}_{R^{-1}}\cdot R^{-1}\bm{t}}e^{i\bm{k}\cdot R^{-1}\bm{t}}=e^{i(-R\bm{\kappa}_{R^{-1}}+R\bm{k})\cdot \bm{t}}.
\end{equation}
Using (\ref{eq:kappa}), we see that $R\bm{\kappa}_{R^{-1}} + \bm{\kappa}_R\in\hat{L}$, then, Eq.~(\ref{21}) becomes
\begin{equation}
	e^{i\bm{k}\cdot\bm{t}}\mapsto e^{i(R\bm{k}+\bm{\kappa}_{R})\cdot \bm{t}}.
\end{equation}
This confirms Eq.~(\ref{eq:R_action}) for the action of $R\in P$ on momentum space. And Eq.~\eqref{eq:kappa} indicates $\bm{\kappa}_R$ has the same structure as real space fractional lattice translations, except that it is in the momentum space. That is, $\bm{\kappa}_R$ can be interpreted as the fractional reciprocal lattice translation associated to the point group element $R$ in momentum space. For the ordinary representation of $G$, all $\bm\kappa_R=0$, so this recovers the conventional wisdom that space group actions on momentum space are symmorphic. However, our discussion above explicitly demonstrates that
for projective representations of $G$, the action will be nonsymmorphic in momentum space for nontrivial $\bm\kappa_R$.

Based on the above discussion, one can construct all kinds of momentum-space nonsymmorphic space groups (denoted as $k$-NSGs). For any given $k$-NSG with reciprocal lattice $\hat L$, point group $P$, and crystal class $\hat D$, 
the set of $\bm\kappa_R$ is also fixed. One first finds the real space lattice $L$ from $\hat L$ (simply via the Fourier transform). $P$ and $L$ determines the real space crystal class $D$.
Then one can construct the real-space (symmorphic) space group $G=L\rtimes_D P$. The required $k$-NSG just corresponds to the projective representation of $G$ with the multipliers $\nu$ given by Eq.~\eqref{eq:multiplier}.


Consider the simple example of $k$-NSG $Pg$, a 2D wallpaper group. Besides translations along two orthogonal directions (labeled as $x$ and $y$),
there is only one glide mirror element 
\begin{equation}
  \mathcal{M}_x:~(k_x,k_y)\mapsto (-k_x,k_y+G_y/2),
\end{equation}
where reciprocal lattice constant along $y$. To realize this $k$-NSG, following the recipe above, we find the corresponding real space group $G$ can be chosen as $Pm$, which just contains translations generated by primitive ones $L_x$ and $L_y$, as well as (symmorphic) mirror $M_x$. 

From Eq.~(\ref{eq:R_action}), one finds that $\bm\kappa_{M_x}=\bm G_y/2$. Using Eq.~(\ref{eq:multiplier}), the corresponding factor system can be determined. For example, one can see that
\begin{equation}\label{eq:LM_factor}
	\nu(M_x,L_y)=-1, \quad \nu(L_y,M_x)=1.
\end{equation}
Thus, $Pg$ is realized as a $k$-NSG by the projective representation of $Pm$ with this factor system $\nu$. 

To explicitly see the nonsymmorphic character of $\rho(M_x)$ in this projective representation, one notes that from 
Eq.~(\ref{eq:LM_factor}), we have 
\begin{equation}\label{eq:p_LyMx}
	\rho(M_x)\rho(L_y)=-\rho(L_y)\rho(M_x).
\end{equation}
In momentum space, the translation operator $\rho(L_y)$ is decomposed into ${k}$-components, $\rho_{\bm{k}}(L_y)=e^{ik_yb}$ with $b$ the lattice constant along $y$. Then, from \eqref{eq:p_LyMx}, we see 
\begin{equation}
  [\rho(M_x)]^\dagger e^{ik_yb}\rho(M_x)=-e^{ik_yb}=e^{i(k_y+G_y/2)b}.
\end{equation}
This clearly shows that under the projective representation, $\rho(M_x)$ gives the glide mirror operation $\mathcal M_x$ acting on momentum space.




In genral, although not all projective representations of space groups can be realized by
lattice models with gauge fluxes, those corresponding to $k$-NSGs can always be realized in this way. This can be inferred from the particular form of Eq.~\eqref{eq:multiplier}. It only modifies the algebraic relations between translations and point-group symmetries, namely, modifying the operation of the point group on the real-space lattice with additional phase factors [see Eq.~\eqref{eq:Conjugation}]. These phase
factors can be realized by gauge fluxes, resembling the
Aharonov–Bohm effect. More technical details can be found in Ref. ~\cite{chen2022brillouin}.

\begin{table*}[!t]
    \centering
    \renewcommand\arraystretch{1.2}
	\begin{tabular}{|c|c|c|c|c|c|c|c|c|}
		\hline
		$\alpha$&Platycosms $\M^{\alpha}$&Ori.& $B^{\alpha}$&Crystal class &Fractional Translations &$\widetilde{K}(\mathcal{M}^\alpha)$ & $S^{\alpha}$&PCA relations  \\
		\hline
		0&Cubical torocosm&Y&P1&1P  &None&$\Z^3$ & P1&None\\
		\hline
		1&First amphicosm&N&Pc&mP   &$\bm{\kappa}_{M_x}=\frac{1}{2}G_z$&$\Z_2\oplus\Z$& Pm&$M_xL_zM_x^{-1}L_z^{-1}=-1$\\
		\hline
		2&Second amphicosm&N&Cc&mC &$\bm{\kappa}_{M}=\frac{1}{2}G_z$&$\Z$&Cm&$ML_zM^{-1}L_z^{-1}=-1$\\
		\hline
		3&First amphidicosm&N&Pca$2_1$&mm2P &$\begin{array}{c}\bm{\kappa}_{M_x}=\frac{1}{2}G_z\\\bm{\kappa}_{M_y}=\frac{1}{2}G_x\end{array}$&$\Z_2^2$& Pmm2&$\begin{array}{c}
			M_xL_zM_x^{-1}L_z^{-1}=-1   \\
			M_yL_xM_y^{-1}L_x^{-1}=-1
		\end{array}$  \\
		\hline
		4&Second amphidicosm&N&Pna$2_1$&mm2P&$\begin{array}{c}\bm{\kappa}_{M_x}=\frac{1}{2}(G_y+G_z)\\\bm{\kappa}_{M_y}=\frac{1}{2}G_x \end{array}$ &$\Z_4$& Pmm2&$\begin{array}{c}
			M_xL_yM_x^{-1}L_y^{-1}=-1 \\
			M_xL_zM_x^{-1}L_z^{-1}=-1\\
			M_yL_xM_y^{-1}L_x^{-1}=-1
		\end{array}$\\
		\hline
		5&Dicosm&Y&P$2_1$&2P  &$\bm{\kappa}_{R_2}=\frac{1}{2}G_z$&$\Z_2^2\oplus\Z$& P2&$R_2L_zR_2^{-1}L_z^{-1}=-1$\\
		\hline
		6&Tricosm&Y&P$3_1$&3P  &$\bm{\kappa}_{R_3}=\frac{1}{3}G_z$&$\Z_3\oplus\Z$& P3&$R_3L_zR_3^{-1}L_z^{-1}=e^{i\frac{2\pi}{3}}$\\
		\hline
		7&Tetracosm&Y&P$4_1$&4P   &$\bm{\kappa}_{R_4}=\frac{1}{4}G_z$&$\Z_2\oplus\Z$& P4&$R_4L_zR_4^{-1}L_z^{-1}=e^{i\frac{\pi}{2}}$\\
		\hline
		8&Hexacosm&Y&P$6_1$&6P&$\bm{\kappa}_{R_6}=\frac{1}{6}G_z$&$\Z$&P6&$R_6L_zR_6^{-1}L_z^{-1}=e^{i\frac{\pi}{3}}$\\
		\hline
		9&Didicosm&Y&P$2_12_12_1$&222P &$\begin{array}{c} \bm{\kappa}_{R_z}=\frac{1}{2}(G_y+G_z)\\\bm{\kappa}_{R_y}=\frac{1}{2}(G_x+G_y) \end{array}$ &$\Z_4^2$& P222&$\begin{array}{c}
			(R_zL_y)^2=-1\\
			(R_yL_x)^2=-1  \\
			R_zL_zR_z^{-1}L_z^{-1}=-1\\
			R_yL_xR_z^{-1}L_x^{-1}=-1
		\end{array}$\\
		\hline 		
	\end{tabular}
    \caption{Table of Brillouin platycosms. The first two columns list all ten platycosms $\mathcal{M}^{\alpha}$ and their labels $\alpha = 0, \dots, 9$. Among them, orientable ones are marked with 'Y' in the third column, while non-orientable ones are marked with 'N'. The fourth column lists the corresponding Bieberbach groups $B^{\alpha}$. The following two columns provide their crystal classes and momentum space fractional translations. In the seventh column, we present the reduced K groups $\tilde{K}(\mathcal{M}^{\alpha})$ of the Brillouin platycosms $\mathcal{M}^{\alpha}$. The eighth column lists the corresponding real-space symmorphic groups $S^{\alpha}$, while the last column contains the projective crystal algebra relations of $S^{\alpha}$.}
    \label{table1}
\end{table*}

\section{Fundamental domain and topological phases}
An important consequence of the appearance of $k$-NSG is the dramatic change of fundamental domain of momentum space. 

Still taking $k$-NSG $Pg$ as an example, under glide mirror operation $\mathcal M_x$, the two $k$ points 
$(k_x,k_y)$ and $(-k_x,k_y+\pi)$ are interconnected, allowing the Brillouin zone to be decomposed into two parts $\tau_{1/2}$ and $\bar{\tau}_{1/2}$, as depicted in Fig.~\ref{fig_klein}(a). We may choose the fundamental domain to be $\tau_{1/2}$. Importantly, $\tau_{1/2}$ is a closed 2D manifold having topology of a Klein bottle $K^2$. This can be seen from Fig.~\ref{fig_klein}(b). Note that while the left and the right edges of  $\tau_{1/2}$ (the red lines) can be directly identified, the top and the bottom edges must be glued after a twist, because these two edges have opposite orientations due to $\mathcal M_x$. 
Thus, under projective crystal symmetry, we can obtain an unprecedented type of momentum space fundamental domain: The topology of a torus Brillouin zone is now replaced by a Klein bottle. 

This change in fundamental domain topology also alters the topological classification of band structures. 
A band structure can be viewed as a vector bundle over the momentum space manifold. 
The classification is simplified for the case of nonsymmorphic symmetry in momentum space, because the action of nonsymmorphic symmetry is free, meaning that it has no fixed point.
Mathematically, when a structure group $G$ acts freely on a base space $X$, the equivariant bundle becomes equivalent to a bundle over the quotient space $X/G$. For the current case, it simply means all the information including topology is fully captured by the band structure over the fundamental domain $\tau_{1/2}$.

In Ref.~\cite{chen2022brillouin}, it was shown that the topological classification of insulating phases over the Brillouin Klein bottle is characterized by a $\mathbb{Z}_2$ invariant, which can be computed from
\begin{equation}
    \nu=\frac{1}{2\pi}\int_{\tau_{1/2}}\mathcal{F}+\frac{1}{\pi}\gamma(0)~\mathrm{mod}~2.
\end{equation}
Here, $\mathcal{F}$ denotes the Berry curvature of valence bands, and $\gamma(0)$ is the Berry phase over the path $S^1_{+}$ (see Fig.~\ref{fig_klein}(a)). This $\mathbb{Z}_2$ classification is based on the equivariant K-theory classification of vector bundles, specifically, $\tilde{K}(K^2)=\mathbb{Z}_2$. 

For the nontrivial phase with $\nu=1$, the physical consequence is the existence of gapless edge bands 
under open boundary conditions in $x$ direction.
Notably, 
the edge states here are protected by the projective algebra of $M_x$ and $L_y$ symmetries, thereby requiring the preservation of translational symmetry along $y$.

Since a fundamental domain is a quotient space obtained from the flat momentum space, it should be both flat and compact. 
In 2D, there are only two kinds of flat and compact manifolds, i.e., the torus $T^2$ and the Klein bottle $K^2$. 
The discussion above shows both kinds can be realized as fundamental domains of momentum space.

This idea can be naturally generalized to three dimensions~\cite{prl136601}. In 3D, there are ten flat and compact manifolds, known as the ten platycosms, as listed in Table~\ref{table1}. While the zeroth platycosm $\mathcal{M}^{0}$ is just the three torus $T^3$, the other nine platycosms have more intricate topologies. Among them, four manifolds, namely $\alpha=1,2,3,4$, are non-orientable, as translation along certain directions induces a mirror reflection. They all feature non-trivial 1D cycles with finite orders. Particularly, all elementary 1D cycles in the didicosm are fourfold, namely,
traveling along any cycle four times necessarily leads to
a contractible cycle, which inspired the popular science
fiction story, \textit{Didicosm}~\cite{didicosm_novel} 

We find that they can all be realized as momentum-space fundamental domains
under certain 3D crystal groups.  These groups act freely on $\mathbb{R}^3$ (momentum space) and they are known as the 
ten Bieberbach groups $B^{\alpha}$ with $\alpha=0,1,\dots,9$. Bieberbach groups are torsion-free and admit a maximal free abelian subgroup of finite index.  The ten platycosms $\mathcal{M}^{\alpha}$ just correspond to the ten quotient manifolds
%
%
%
\begin{equation}
    \mathcal{M}^{\alpha}=\mathbb{R}^3/B^{\alpha}.
\end{equation}

Following the construction outlined above, each Brillouin platycosm $\mathcal{M}^{\alpha}$ can be generated from projective representation of some real-space symmorphic group $S^{\alpha}$ (see Table~\ref{table1}). Within projective crystal symmetry framework, these ten platycosms constitute all possible types of fundamental domains for 3D crystals, since every 3D crystallographic group has one of the ten platycosms reduced from the maximal Bieberbach subgroup as its momentum-space unit.

The nine nontrivial platycosms exhibit distinct topological signatures that generate rich distinctive topological phases, which are listed in Table~\ref{table1}.  By using the method of Atiyah-Hirzebruch spectral sequence, we find an interesting result that the reduced K group of each platycosm is isomorphic to the second cohomology group of the corresponding Bieberbach group:
\begin{equation}
    \tilde{K}(\mathcal{M}^{\alpha})\cong H^2(B^{\alpha},\mathbb{Z}).
\end{equation}
This isomorphism enables systematic construction of complete topological invariants for each class, though the detailed derivations lie beyond our present scope.

In general, transitions between two such 3D topological phases are characterized by a Weyl semimetal phase. This can be attributed to the fact that no crystal symmetry exists at a generic point in the Brillouin platycosm (since the action of the Bieberbach group on momentum space is free). Consequently, crossing points of energy bands are generically twofold degenerate Weyl points. These Weyl semimetals satisfy a generalized Nielsen-Ninomiya theorem~\cite{NIELSEN1981219,PhysRevLett.132.266601}. For orientable Brillouin platycosms, the total chirality numbers of Weyl points over them must equal to zero. On the other hand, for non-orientable ones the total chirality numbers can be any even integers. The latter can be understood by noting that for a non-orientable space, the left and the right handedness have no distinction in the global sense, so topological charges $-1$ and $+1$ are equivalent. In other words, the originally $\mathbb{Z}$-valued topological charge now becomes $\mathbb{Z}_2$-valued.
Recently, these have been demonstrated for the case with Brillouin first amphicosm. The similar scenario may also occur in 2D. For example, when the fundamental domain has the Klein bottle topology (which is also non-orientable) and when the system also possesses chiral symmetry, the $\mathbb{Z}$-value topological charge of a Dirac point will also be reduced to $\mathbb{Z}_2$-valued.

\section{effects of internal symmetry}
Thus far, we have considered spatial symmetries, like translation and rotation, which operate on real space. 
For physical systems, there may also exist internal symmetries, which do not act spatially but strongly affect the system's properties. In condensed matters, the most well known ones are the time-reversal symmetry ($T$), the particle-hole symmetry, and the chiral/sublattice symmetry. They form the basis of the tenfold Altland-Zirnbauer symmetry classes. 

%
%

Combining internal symmetries with projective crystal symmetries may produce rich and fascinating results. 
Several interesting cases have been reported in recent years. 
The first example occurs in systems with projective $PT$ symmetry, where $P$ denotes spatial inversion~\cite{PhysRevLett.126.196402}. Ordinarily, for spin-$s$ particles, these operators obey $T^2 = (-1)^{2s}$ and $P^2 = 1$. They commute with each other, i.e., $[P,T]=0$, and therefore
\begin{equation}\label{PT1}
    (PT)^2=(-1)^{2s}.
\end{equation}

Nontrivial gauge flux can projectively alter this algebra. Under specific gauge flux configurations, the proper inversion $\mathcal{P}$, i.e., the projective representation of $P$, becomes the composition of inversion operation $P$ and a gauge transformation $\mathrm{G}$. We can write
\begin{equation}
    \mathcal{P}=\mathrm{G}P.
\end{equation}

Specifically, $\mathrm{G}$ preserves time-reversal symmetry $T$ and satisfies $\mathrm{G}^2 = 1$. Furthermore, if $P$ reverses the gauge transformations, i.e., $\mathrm{G}$ anti-commutes with $P$, then we have the following relations:
\begin{equation}
    [\mathrm{G},T]=0,~\{\mathrm{G},P\}=0,~\mathrm{G}^2=1.
\end{equation}
This implies $\mathcal{P}^2 = (\mathrm{G}P)^2 = -1$. Within the projective representation framework, these relations indicate a nontrivial factor system for the point group generated by the inversion symmetry. Consequently, the projective spacetime inversion symmetry $\mathcal{P}T$ obeys the modified algebra:
\begin{equation}\label{PT2}
    (\mathcal{P}T)^2=\mathcal{P}^2T^2=(-1)^{2s+1}.
\end{equation}
This is a remarkable result, since comparing (\ref{PT1}) and (\ref{PT2}) shows the fundamental symmetry algebra is exchanged for spinless and spinful systems. It follows that their topological classifications are also exchanged.

Another example is the spinless mirror Chern insulator~\cite{PhysRevB.108.205126}. 
It was previously believed that a mirror Chern insulator must require spin-orbit coupling, since time-reversal symmetry for spinless systems forbids nontrivial mirror Chern number~\cite{hsieh2012topological,tanaka2012experimental,teo2008surface}. However, when projectively represented, the algebra between mirror $\mathcal M$ and time-reversal symmetry $T$ may satisfy
\begin{equation}
    [T,\mathcal M]=0,~\mathcal M^2=-1.
\end{equation}
This modified projective symmetry algebra actually allows to establish the mirror Chern insulator in spinless systems~\cite{PhysRevB.108.205126}. The $\mathcal M^2=-1$ condition generates Kramers-like degeneracy, while the $[T,\mathcal M]=0$ relation preserves time-reversal-protected band crossings. This also demonstrates that the essential constraint on mirror Chern insulator originates from the symmetry operator algebra, rather than spin-orbit coupling.

As mentioned, chiral/sublattice symmetry is also an important internal symmetry often considered in topological band theory. In literature, the sublattice symmetry $S$ and the chiral symmetry $\Pi$ usually are not distinguished~\cite{chiu2016classification,altland1997nonstandard,heinzner2005symmetry,schnyder2008classification}. However, the projective algebra between sublattice and translation symmetries reveals an essential difference~\cite{xiao2024revealing}. This is because $S$ inherently possesses spatial dependence through its bipartite lattice partitioning, a feature absent in the chiral symmetry. This spatial nature actually allows classifying $S$ into two classes based on its commutativity with translation operators $\mathcal L_i$.

In class I, the sublattice symmetry commutes with all translation operators, i.e., $[\mathcal L_i,S]=0$. Class-I $S$ can be effectively regarded as an internal symmetry in the momentum space, thereby inheriting all topological classifications for chiral symmetry $\Pi$.

Meanwhile, systems in which translation operators exchange sublattices demand a different class-II characterization. For class II $S$, at least one translation operator, say $\mathcal L_x$, anti-commutes with $S$,
\begin{equation}
    \{\mathcal L_x,S\}=0.
\end{equation}
This is because $\mathcal L_x$ inverts the sign assigned by $S$ on each sublattice. Through projective representation analysis, we uncover a momentum-shifting action of $S$:
\begin{equation}
    S:k_x\mapsto k_x+G_x/2,
\end{equation}
which fundamentally alters the topological classification. Compared to the topological classifications for class AIII or class-I sublattice symmetry, there are two significant differences: i) The classification for class-II $S$ symmetry is nontrivial in even dimensions, while it is nontrivial in odd dimensions for class AIII; ii) The nontrivial classification shifts from $\mathbb{Z}$ to $\mathbb{Z}_2$ under class-II $S$ constraints.

\begin{figure}[!t]
    \centering
    \includegraphics[width=\linewidth]{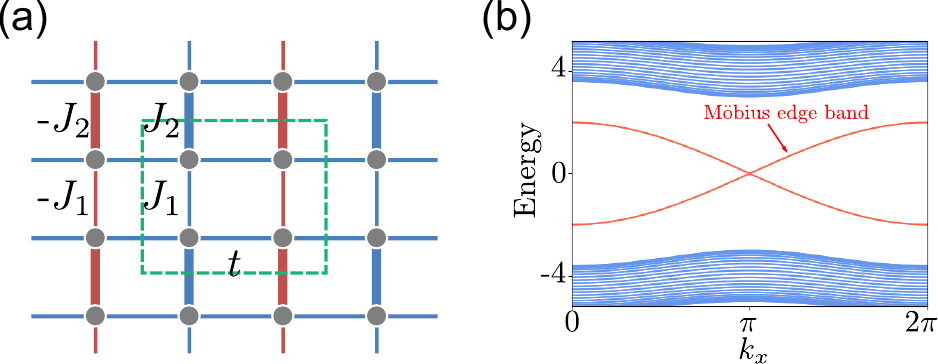}
    \caption{(a) Lattice model of the M{\"o}bius topological insulator.  Hopping amplitudes's negative and positive signs are marked in red and blue, respectively, and their magnitudes are indicated by the thickness of the lines. The green-dashed rectangle outlines a unit cell containing four 
    sites. (b) The band structure of the model under a slab geometry. The boundaries are opened normal to the $y$ direction, and therefore the edges are long the $x$ direction. Parameter values are set as $t=1,J_1=1$, and $J_2=4$.}
    \label{fig_mobius}
\end{figure}

The M{\"o}bius topological insulator~\cite{zhao2020z} mentioned above actually exemplifies the class-II sublattice symmetry. Consider a 2D lattice model [see Fig.~\ref{fig_mobius}(a)] with Bloch Hamiltonian
\begin{equation}
\begin{split}
    \mathcal{H}(\bm{k})=&t(1+\cos k_x)\Gamma^1+t\sin k_x\Gamma^2 \\
    &+(J_1+J_2\cos k_y)\Gamma^3+J_2\sin k_y\Gamma^4,
\end{split}
\end{equation}
where $\Gamma^1=\tau_0\otimes\sigma_1,\Gamma^2=\tau_0\otimes\sigma_2,\Gamma^3=\tau_1\otimes\sigma_3,\Gamma^4=\tau_2\otimes\sigma_3$.  Here $\tau_i$ and $\sigma_i$ for $i=1,2,3$ are two sets of the standard Pauli matrices where $\tau_0=\sigma_0$ is the $2\times 2$ identity matrix. This model features a four-site unit cell with hopping amplitudes $t,J_1,J_2>0$ as labeled in Fig.~\ref{fig_mobius}(a), where the signs of hopping amplitudes are distinguished by different colors. The intra- and inter-cell hoppings along $x$ are both $t$, whereas when $J_1 \neq J_2$, the system exhibits dimerization along $y$.
One can show that this model possesses a sublattice symmetry $S$ which anticommutes with $\mathcal L_x$~\cite{zhao2020z}. Hence, this algebraic constraint reduces the topological classification to $\mathbb{Z}_2$, distinct from conventional 2D chiral symmetric systems~\cite{chiu2016classification,schnyder2008classification}. The Möbius topology can be understood as a direct consequence of the projective symmetry action $S: k_x \mapsto k_x+\pi$, which effectively identifies $k_x$ and $k_x+\pi$ with opposite sublattice parity. In the M{\"o}bius insulator phase, there is a pair of zero models at $k_x=\pi$ for an edge perpendicular to $y$, as illustrated in Fig.~\ref{fig_mobius}(b). The two edge bands are connected to each other, forming a M{\"o}bius twist over the edge Brillouin zone~\cite{zhao2020z}. \\
\section{Outlook}
We have presented the core principles behind the recent advancements in the emerging field of projective crystal symmetry. These foundational ideas have already led to numerous ground-breaking discoveries, expanded our understanding of novel states of matter, and promise significant potential for various applications. Nevertheless, this field remains in its early stages, with many important and promising research endeavors awaiting further investigation.

First, the studies so far are demonstrated on a selective collection of symmorphic crystal space groups,  
therefore, a systematic study of projective representations of all space group is an important task to be undertaken in the future.
Nonsymmorphic space groups could have comparatively more complicated projective algebra structures. Although
the nonsymmorphic character in real space should not change the key conclusions discussed above in any qualitative way, they could enrich the topological band structure features and need to be investigated.

Second, the topological classification for general projective crystal symmetries, including all momentum-space nonsymmorphic symmetries, remains an open challenge. Detailed studies of specific systems have already revealed several novel topological states~\cite{PhysRevLett.132.213801,PhysRevApplied.21.044002}. However, we still need to develop a systematic approach and to apply the technique for studying all the cases.


Third, the framework should be further extended by systematically incorporating internal symmetries, e.g., within the tenfold Altland-Zirnbauer symmetry classes. Although some intriguing results have been obtained in this area, a comprehensive theory has yet to be fully developed.

Fourth, the physical consequences of projective crystal symmetry should be further explored. 
Regarding the many novel topological insulating or semimetal states enabled by projective symmetries, 
what will be their bulk-boundary correspondence? How will their unique topological band features manifest in any physical properties? These important questions need to be answered.

Finally, more physical realizations of projective crystal symmetry need to be explored. So far, 
the
physical realization is mainly in artificial crystals with engineered gauge flux configurations. Although this 
has already produced great success, it is still desirable to find realization in solid material systems. 
It is noteworthy that recently, two classes of solid systems have been proposed as candidates. One class includes magnets that are invariant under non-collinear spin space groups~\cite{xiao2024spin}, and the other includes twisted Moiré layered material systems~\cite{cualuguaru2024new}. The momentum-space nonsymmorphic symmetries could appear in these systems.  These condensed matter systems can significantly broaden the scope of applications for the theories introduced in this article.

\begin{acknowledgments}
	This work was supported by the GRF of Hong Kong (No.~17301224), the NSFC/RGC JRS grant (No.~12161160315), and
the HK PolyU start-up fund (P0057929). 
\end{acknowledgments}

\vskip 12pt
\appendix

\bibliographystyle{apsrev4-2}
\bibliography{references}
\end{document}